\documentclass[11pt]{article}

\usepackage{amsfonts}
\usepackage{amsmath}
\usepackage{amssymb}
\usepackage{indentfirst}
\usepackage{graphicx}

\begin{document}

\title{Pure Lovelock gravity and Chern-Simons theory}
\author{P. K. Concha$^{1,2}$\thanks{%
patillusion@gmail.com},\thinspace\ R. Durka$^{3}$\thanks{%
remigiuszdurka@gmail.com},\thinspace\ C. Inostroza$^{4}$\thanks{%
c.inostroza@gmail.com}, N. Merino$^{3}$\thanks{%
nemerino@gmail.com},\thinspace\ E. K. Rodr\'{\i}guez$^{1,2}$\thanks{%
everodriguezd@gmail.com}, \\
{\small $^{1}$\textit{Departamento de Ciencias, Facultad de Artes Liberales
y }}\\
{\small \textit{Facultad de Ingenier\'{\i}a y Ciencias, Universidad Adolfo Ib%
\'{a}\~{n}ez,}}\\
{\small Av. Padre Hurtado 750, Vi\~{n}a del Mar, Chile}\\
{\small $^{2}$\textit{Instituto de Ciencias F\'{\i}sicas y Matem\'{a}ticas,
Universidad Austral de Chile,}}\\
{\small Casilla 567, Valdivia, Chile}\\
{\small $^{3}$\textit{Instituto de F\'{\i}sica, Pontificia Universidad Cat%
\'{o}lica de Valpara\'{\i}so,} }\\
{\small Casilla 4059, Valpara\'{\i}so, Chile}\\
{\small $^{4}$\textit{Departamento de F\'{\i}sica, Universidad de Concepci%
\'{o}n,}}\\
{\small Casilla 160-C, Concepci\'{o}n, Chile}}
\maketitle

\begin{abstract}
We explore the possibility of finding Pure Lovelock gravity as a particular limit of a Chern-Simons action for a specific expansion of the AdS algebra in odd dimensions. We derive in details this relation at the level of the action in five and seven dimensions. We provide a general result for higher dimensions and discuss some issues arising from the obtained dynamics.
\end{abstract}

\vspace{-13.5cm}

\begin{flushright}
{\footnotesize UAI-PHY-16/04}
\end{flushright}

\vspace{11.5cm}

\section{Introduction}

Lanczos-Lovelock theory \cite{Lovelock:1971yv,Lanczos} is the most natural
generalization of the gravity theory in $D-$dimensions that leads to the
second order field equations. Its action is built as a polynomial in the
Riemann curvature with each term carrying an arbitrary coupling constant.
As recently discussed in Ref.~\cite{Dadhich:2015ivt}, the higher curvature terms can generate sectors in the space of solutions carrying different number of degrees of freedom. In particular, there might be degenerate sectors with no degrees of freedom and in that case the metric is not completely determined by the field equations, i.e., some components remain arbitrary. An explicit example was first provided in Refs.~\cite{Wheeler:1985nh,Wheeler:1985qd} where it was shown that the metric component $g_{tt}$ of the static spherically symmetric solution stays arbitrary when the coupling constants cause the theory to admit a non-unique degenerate vacuum. In fact, metrics with the undetermined components were reported not only in the static case \cite{Zegers:2005vx} but also when the time dependent metric functions \cite{Charmousis:2002rc,Dotti:2007az,Maeda:2011ii} are considered or when torsional degrees of freedom are taken into account to find a charged solution \cite{Giribet}.

The degeneracy problem mentioned above can be avoided in the static spherically symmetric sector for particular choices of the coupling constants in the Lovelock theory. The simplest case corresponds to the Einstein-Hilbert (EH) action with or without the cosmological constant term. The next one is given by the action containing the Einstein-Hilbert and Gauss-Bonnet terms in $D$ dimensions \cite{Boulware:1985wk}. Another possibility is to choose the coefficients as in the Chern-Simons (CS) or Born-Infeld (BI) theories. Recently in Refs.~\cite{Cai:2006pq,Dadhich:2012ma} the Pure Lovelock (PL)
theory was proposed as another way of fixing the constants such that the
static metric is free of the degeneracy. Indeed, the PL theory has non-degenerate vacua in even dimensions and admits a unique non-degenerate (A)dS vacuum in the odd-dimensional case. In the first order formalism with the
vielbein $e^{a}$ and the spin connection $\omega^{ab}$, the considered
action contains only two terms instead of the full Lovelock series,
\begin{equation}
I^{D}_{\text{PL}}=\int\left( \alpha_{0}\mathcal{L}_{0}+\alpha_{p}\mathcal{L}%
_{p}\right)\, ,  \label{PLgeneral}
\end{equation}
where
\begin{align}
\mathcal{L}_{0} & =\epsilon_{a_{1}\ldots a_{D}}\,e^{a_{1}}\cdots e^{a_{D}}\,,
\\
\mathcal{L}_{p} & =\epsilon_{a_{1}\ldots a_{D}}\,R^{a_{1}a_{2}}\cdots
R^{a_{2p-1}a_{2p}}e^{a_{2p+1}}\cdots e^{a_{D}}\,.
\end{align}
As usual, the curvature and the torsion are defined as $R^{ab}=d\omega
^{ab}+\omega_{\ c}^{a}\ \omega^{cb}$ and $T^{a}=de^{a}+\omega_{\ b}^{a}\
e^{b}$, where wedge product is assumed and the Lorentz indices $a,b$ are
running from $1$ to $D$. This action naturally carries torsional degrees of
freedom and usually, when the study of solutions is considered (see Ref.~%
\cite{Exirifard:2007da}), the vanishing of torsion is assumed.

The new coefficients, in terms of the cosmological constant related with the
(A)dS radius $\ell$
\begin{equation}
\Lambda=\dfrac{\left( \mp1\right) ^{p}\left( D-1\right) !}{2\left(
D-2p-1\right) !\ell^{2p}}\,,  \label{cosmologicalconstant}
\end{equation}
and the gravitational constant $\kappa$, will be given by%
\begin{equation}
\alpha_{0}=-\frac{2\Lambda\kappa}{D!}=-\dfrac{\left( \mp1\right) ^{p}\kappa
}{D\left( D-2p-1\right) !\ell^{2p}}\,,\qquad\alpha_{p}=\frac{\kappa }{(D-2p)!%
}\,.  \label{cts}
\end{equation}
As was pointed out in Ref.~\cite{Aranguiz:2015voa}, the (A)dS space in the
PL theory is not directly related to the sign of a cosmological constant
like in General Relativity. Indeed, the definition of (A)dS space is due to
the sign of curvature $R^{ab}=\mp\frac{1}{\ell^{2}}e^{a}e^{b}$, and not of
the explicit $\Lambda$, which now has the dimension of the $length^{-2p}$
(at the same time the dimension of the gravitational constant $\kappa$ is $%
length^{2p-D}$).

It was shown in Ref.~\cite{Dadhich:2012ma,Dadhich:2012eg} that the
corresponding PL black holes solutions have the remarkable property of being
asymptotically indistinguishable from the AdS-Schwarzschild ones. In
addition, for the maximal power $p=\left[\frac{D-1}{2}\right]$ they exhibit
a peculiar thermodynamical behavior. From the dynamical point of view, the
nonlinear character of this theory is responsible for the presence of
sectors in the space of solutions carrying different number of degrees of
freedom. The Hamiltonian analysis performed in Ref.~\cite{Dadhich:2015ivt}
shows that the maximum possible number is the same as in the EH case.

It should be pointed out that a supersymmetric version of the general
Lovelock theory is unknown, except for the EH Lagrangian and two very
special cases in odd dimensions corresponding to the CS actions for the AdS
and Poincar\'{e} symmetries. The existence of supergravity for the
Einstein-Hilbert term coupled to a Lovelock quadratic curvature term in 5D
was discussed in \cite{Deser:2011zk}, nevertheless, the explicit form
remains unknown. Regarding the features of the PL theory mentioned earlier,
it would be interesting to explore the possibility of finding its
supersymmetric form. This is not a trivial task because of the lack of
guidance principle with the kind of terms that must be considered to ensure
the invariance under supersymmetric transformations, which also are not
fully clear. However, if one could find the PL action as some kind of limit
of a CS or BI theory for a special symmetry, then that result might be
useful to obtain the supersymmetric Pure Lovelock (sPL) theory. Something
similar has been done using the Lie algebra expansion methods \cite%
{AIPV,Sexp} to obtain General Relativity from the Chern-Simons theory in odd
dimension \cite{Zan, GRCS,CPRS2}. The same has also been achieved in even
dimensions with the BI actions \cite{CPRS1}.

We will not deal with the full problem here, but only focus on the bosonic
part in odd dimensions. Thus, as a first step, in Section 2 we give a brief
review of the so called Maxwell like algebras and the $S$-expansion method
\cite{Sexp} used for their construction, as it provides necessary building
blocks for the actions. We choose one specific family of algebras that seems
to have the best chance to establish the desired relation between PL and CS.
In Section 3 we explicitly construct the CS action in $5D$ and show how it
relates with the PL in that context. For completeness, in Section 4 we
present similar construction for $7D$, along with the extension of the result to $(2n+1)$-dimensions. Finally, after dealing with the number of obstacles (assuring the proper relative sign between the terms, overconstraining superposition of the field equations) we propose specific configuration that reproduces the correct PL dynamics.

\section{$\mathfrak{C}_{m}$ algebras and $S$-expansion}

Our task requires finding the symmetry for which the cosmological constant
term and the desired Lovelock term effectively end up in one sector of the
invariant tensor and unwanted terms in another. The particular symmetries
offering such a feature originate from $\mathfrak{so}(D-1,2)\oplus \mathfrak{%
so}(D-1,1)$ algebra (so called AdS-Lorentz introduced in Refs.~\cite%
{Sorokas,Sorokas2,DKGS,SS}), where we equip the standard set $%
\{J_{ab},P_{a}\}$ with a new generator $Z_{ab}$. The abelian semigroup
expansion procedure (also referred as the $S$-expansion) \cite{Sexp} has
generalized it to the whole family, which we will denote as $\mathfrak{C}%
_{m} $, following notation proposed in \cite{CDMR} and where the subindex
refers to $(m-1)$ types of generators. Although we extend the AdS algebra,
it is worth to mention that for $m>4$ the new algebras do not contain it as
a subalgebra. The $\mathfrak{C}_{m}$, as well as $\mathfrak{B} _{m}$ and $%
\mathfrak{D}_{m}$ families, introduced in Refs.~\cite{GRCS, SS, CDMR}, can
be regarded as generalizations of the original Maxwell algebra \cite%
{Schrader, BCR} and thus they are all referred as Maxwell like algebras. We
focus on the $\mathfrak{C}_{m}$ family because for its first representatives
we are going to consider in 5D one can arrive to the field equations
resembling, at least at the first sight, the wanted scheme (effectively
giving maximal PL like equation $RR+eeee=0$), whereas $\mathfrak{B}_{m}$
(producing $RR=0$ and $Ree=0$) and $\mathfrak{D}_{m}$ (producing $RR=0$ and $%
Ree+eeee=0$) explicitly exclude the maximal PL construction.

As we will see below, the $S$-expansion method is not only producing new
algebras, but it will also deliver the corresponding invariant tensors
needed to construct the actions. This makes it a very useful and important
tool to relate different symmetries and (super)gravity theories \cite{GRCS,
CPRS3, DFIMRSV, BDgrav, CR1, CR2, CFRS}. For example, it was shown in Ref.~%
\cite{SS} that the cosmological constant term can arise from the Born-Infeld
gravity theory based on $\mathfrak{C}_{m}$ (see also \cite{AKL}). Some
further applications of the supersymmetric extension of $\mathfrak{C}_{m}$
can be found in Refs.~\cite{FISV, CRS, DGS}.

Following Ref.~\cite{SS}, in this section we will review the explicit
derivation of the $\mathfrak{C}_{m\geq3}$ algebras from AdS by using a
particular choice of a semigroup. The considered procedure starts from the
decomposition of the original algebra $\mathfrak{g}$ into subspaces,
\begin{equation*}
\mathfrak{g}=\mathfrak{so}\left( D-1,2\right) =\mathfrak{so}\left(
D-1,1\right) \oplus\frac{\mathfrak{so}\left( D-1,2\right) }{\mathfrak{so}%
\left( D-1,1\right) }=V_{0}\oplus V_{1}\,,
\end{equation*}
where $V_{0}$ is spanned by the Lorentz generator $\tilde{J}_{ab}$ and $%
V_{1} $ by the AdS translation generator $\tilde{P}_{a}$. They satisfy the
commutation relations
\begin{align}
\left[ \tilde{J}_{ab},\tilde{J}_{cd}\right] & =\eta_{cb}\tilde{J}%
_{ad}-\eta_{ca}\tilde{J}_{bd}-\eta_{db}\tilde{J}_{ac}+\eta_{da}\tilde{J}%
_{bc}\,,  \notag  \label{AdS1} \\
\left[ \tilde{J}_{ab},\tilde{P}_{c}\right] & =\eta_{cb}\tilde{P}%
_{a}-\eta_{ca}\tilde{P}_{b}\,,  \notag \\
\left[ \tilde{P}_{a},\tilde{P}_{b}\right] & =\tilde{J}_{ab}\,.
\end{align}
The subspace structure may be then written as
\begin{equation}
\left[ V_{0},V_{0}\right] \subset V_{0}\,,\qquad\left[ V_{0},V_{1}\right]
\subset V_{1}\,,\qquad\left[ V_{1},V_{1}\right] \subset V_{0}\,.  \label{eqa}
\end{equation}
We skip the two first representatives, namely $\mathfrak{C}_{3}$ and $%
\mathfrak{C}_{4}$, because as it will be explained in the next section they
do not have necessary features. Thus, for our purposes we proceed with a
detailed derivation of $\mathfrak{C}_{5}$, which we then generalize to an
arbitrary $m$. We start by considering the abelian semigroup $S_{M}^{\left(
3\right) }=\left\{ \lambda_{0},\lambda_{1},\lambda_{2},\lambda_{3}\right\} $
with the multiplication law%
\begin{equation}
\lambda_{\alpha}\lambda_{\beta}=\left\{
\begin{array}{c}
\lambda_{\alpha+\beta},\text{ \ \ \ \ if }\alpha+\beta\leq3\,, \\
\lambda_{\alpha+\beta-4},\text{ \ if }\alpha+\beta>3\,,%
\end{array}
\right.
\end{equation}
and the subset decomposition $S_{M}^{\left( 3\right) }=S_{0}\cup S_{1}\,$,
where $S_{0}=\left\{ \lambda_{0},\lambda_{2}\right\} $ and $S_{1}=\left\{
\lambda_{1},\lambda_{3}\right\} $. This decomposition satisfies
\begin{equation}
S_{0}\cdot S_{0}\subset S_{0}\,,\qquad S_{0}\cdot S_{1}\subset
S_{1}\,,\qquad S_{1}\cdot S_{1}\subset S_{0}\,,
\end{equation}
which is said to be resonant since it has the same structure as (\ref{eqa}).

According to the Theorem IV.2 from Ref.~\cite{Sexp}, we can say that $%
\mathfrak{g}_{R}=W_{0}\oplus W_{1}$ is a resonant subalgebra of $%
S_{M}^{\left( 3\right) }\times\mathfrak{g}$, where
\begin{align}
W_{0} & =\left( S_{0}\times V_{0}\right) =\left\{ \lambda_{0},\lambda
_{2}\right\} \times\left\{ \tilde{J}_{ab}\right\} =\left\{ \lambda _{0}%
\tilde{J}_{ab},\lambda_{2}\tilde{J}_{ab}\right\} \,,  \notag \\
W_{1} & =\left( S_{1}\times V_{1}\right) =\left\{ \lambda_{1},\lambda
_{3}\right\} \times\left\{ \tilde{P}_{a}\right\} =\left\{ \lambda _{1}\tilde{%
P}_{a},\lambda_{3}\tilde{P}_{a}\right\} \,.
\end{align}
The new $\mathfrak{C}_{5}$ algebra is given by the set of generators $%
\left\{ J_{ab},P_{a},Z_{ab},R_{a}\right\} $, defined as
\begin{equation*}
\begin{tabular}{ll}
$J_{ab}=\lambda_{0}\tilde{J}_{ab}\,,$ & $P_{a}=\lambda_{1}\tilde{P}_{a}\,,$
\\
$Z_{ab}=\lambda_{2}\tilde{J}_{ab}\,,$ & $R_{a}=\lambda_{3}\tilde{P}_{a}\,.$%
\end{tabular}%
\end{equation*}
and satisfying
\begin{align}
\left[ P_{a},P_{b}\right] & =Z_{ab}\,,\text{ \ \ \ \ }\left[ J_{ab},P_{c}%
\right] =\eta_{bc}P_{a}-\eta_{ac}P_{b}\,,  \notag \\
\left[ J_{ab,}J_{cd}\right] & =\eta_{bc}J_{ad}-\eta_{ac}J_{bd}+\eta
_{ad}J_{bc}-\eta_{bd}J_{ac}\,,  \notag \\
\left[ J_{ab,}Z_{cd}\right] & =\eta_{bc}Z_{ad}-\eta_{ac}Z_{bd}+\eta
_{ad}Z_{bc}-\eta_{bd}Z_{ac}\,,  \notag \\
\left[ Z_{ab,}Z_{cd}\right] & =\eta_{bc}J_{ad}-\eta_{ac}J_{bd}+\eta
_{ad}J_{bc}-\eta_{bd}J_{ac}\,,  \notag \\
\left[ R_{a},R_{b}\right] & =Z_{ab}\,,\text{ \ \ \ \ }\left[ Z_{ab},P_{c}%
\right] =\eta_{bc}R_{a}-\eta_{ac}R_{b}\,,  \notag \\
\left[ R_{a},P_{b}\right] & =J_{ab}\,,\text{ \ \ \ \ }\left[ J_{ab},R_{c}%
\right] =\eta_{bc}R_{a}-\eta_{ac}R_{b}\,,  \notag \\
\left[ Z_{ab},R_{c}\right] & =\eta_{bc}P_{a}-\eta_{ac}P_{b}\,.  \label{C5}
\end{align}

Analogously we make a transition to the algebras containing arbitrary number
of $(m-1)$ types of generators, with $m\geq3$. Because there is no relation
between spacetime dimension $D$ and $m$, all the algebras $\mathfrak{C}_{m}$
can be used to construct the actions for a given dimension\footnote{%
The dimension of the Lie algebra depends on the spacetime dimension, which
is encoded in the range of the Lorentz indices characterizing the generators.%
}. If the $\mathfrak{C}_{m}$-CS Lagrangian in odd dimensions $D\geq5$ (with $%
m\geq D$) leads to the desired PL action limit, then $\mathfrak{C}_{m+1}$
offers exactly the same result. However, in the last case the theory has
more extra fields as the gauge symmetry is bigger. So, without loosing
generality, we choose the minimal setting and only consider $m$ being odd.
The analysis might be extended for even dimensions $D\geq6$ to obtain PL
from BI gravities based on $\mathfrak{C}_{m}$ with $m$ even, but we will not
deal with that problem here (work in progress).

Thus, as we are restricting ourselves only to the odd values of the $m$
index, we obtain $\mathfrak{C}_{m}$ algebras using the $S_{M}^{\left(
m-2\right) }=\left\{ \lambda _{0},\lambda _{1},\ldots ,\lambda
_{m-2}\right\} $ semigroup with the multiplication law
\begin{equation}
\lambda _{\alpha }\lambda _{\beta }=\left\{
\begin{array}{c}
\lambda _{\alpha +\beta },\text{ \ \ \ \ \ \ \ \ \ \ if }\alpha +\beta \leq
m-2\,, \\
\lambda _{\alpha +\beta -\left( m-1\right) }\,,\text{ \ \ if }\alpha +\beta
>m-2\,,%
\end{array}%
\right.   \label{ML}
\end{equation}%
and the resonant subset decomposition $S_{M}^{\left( m-2\right) }=S_{0}\cup
S_{1}\,$, with%
\begin{align*}
S_{0}& =\left\{ \lambda _{2i}\right\} ,\text{ \ \ \ \ with }i=0,\ldots ,%
\frac{m-3}{2}\,, \\
S_{1}& =\left\{ \lambda _{2i+1}\right\} ,\text{ \ with }i=0,\ldots ,\frac{m-3%
}{2}\,.
\end{align*}%
The new algebra will be spanned by $\left\{ J_{ab,\left( i\right)
},P_{a,\left( i\right) }\right\} $, whose generators are related to the $%
\mathfrak{so}\left( D-1,2\right) $ ones through%
\begin{equation*}
J_{ab,\left( i\right) }=\lambda _{2i}\tilde{J}_{ab}\text{\thinspace
\thinspace\ and\thinspace \thinspace \thinspace }P_{a,\left( i\right)
}=\lambda _{2i+1}\tilde{P}_{a}\,.
\end{equation*}%
General commutation relations will be provided by
\begin{align}
\left[  J_{ab,\left(  i\right)  },J_{cd,\left(  j\right)  }\right]   &
=\eta_{bc}J_{ad,\left(  i+j\right)  \operatorname{mod}\left(  \frac{m-1}%
{2}\right)  }-\eta_{ac}J_{bd,\left(  i+j\right)  \operatorname{mod}\left(
\frac{m-1}{2}\right)  }\nonumber\\
&  +\eta_{ad}J_{bc,\left(  i+j\right)  \operatorname{mod}\left(  \frac{m-1}%
{2}\right)  } -\eta_{bd}J_{ac,\left(  i+j\right)  \operatorname{mod}\left(
\frac{m-1}{2}\right)  }\,,\nonumber\\
\left[  J_{ab,\left(  i\right)  },P_{a,\left(  j\right)  }\right]   &
=\eta_{bc}P_{a,\left(  i+j\right)  \operatorname{mod}\left(  \frac{m-1}%
{2}\right)  }-\eta_{ac}P_{b,\left(  i+j\right)  \operatorname{mod}\left(
\frac{m-1}{2}\right)  }\,,\nonumber\\
\left[  P_{a,\left(  i\right)  },P_{b,\left(  j\right)  }\right]   &
=J_{ab,\left(  i+j+1\right)  \operatorname{mod}\left(  \frac{m-1}{2}\right)
}\,. \label{Cm}%
\end{align}
We can notice that the first non trivial case is given by $\mathfrak{C}_{4}=$
$\mathfrak{so}(D-1,2)\oplus \mathfrak{so}(D-1,1)$, where $\mathfrak{C}_{3}$
trivially corresponds to the original $\mathfrak{so}(D-1,2)$ algebra.

As was mentioned before, quite interestingly, the $S$-expansion method
allows us to express the invariant tensor of the final algebra in terms of
the original one. Based on the definitions of Ref.~\cite{Sexp}, we can show
that the only non-vanishing components of an invariant tensor in $D=2n+1$
spacetime of order n+1 for the $\mathfrak{C}_{m}$ algebra (restricted only
to the odd values of $m$) are given by%
\begin{equation}
\left\langle J_{a_{1}a_{2},\left( i_{1}\right) }\cdots
J_{a_{2n-1}a_{2n},\left( i_{n}\right) }P_{a_{2n+1},\left( i_{n+1}\right)
}\right\rangle =\frac{2^{n}}{n+1}\sigma_{2i+1}\delta_{j\left(
i_{1},i_{2},\ldots,i_{n+1}\right) }^{i}\epsilon_{a_{1}a_{2}\ldots
a_{2n+1}}\,,  \label{invcm}
\end{equation}
where the $\sigma_{j}$'s are arbitrary dimensionless constants and we have
\begin{equation}
j\left( i_{1},i_{2},\ldots,i_{n+1}\right) =\left( i_{1}+i_{2}+\cdots
+i_{n+1}\right) \operatorname{mod}\left( \frac{m-1}{2}\right) \,.  \label{rule}
\end{equation}
With these building blocks at hands we can now turn to the construction of
the corresponding CS actions.

\section{Chern-Simons gravity and Pure Lovelock action in $D=5$}

In the present section we will show how the five-dimensional maximal Pure
Lovelock action (the first non trivial example referred also as the Pure
Gauss-Bonnet \cite{Aranguiz:2015voa, Dadhich:2015ivt}) could be obtained
from a CS gravity theory for a particular choice of the $\mathfrak{C}_{m}$
algebra.

A CS action \cite{Cham1, Cham2} is a quasi invariant (i.e. invariant up to a
boundary term) functional of a gauge connection one-form $A$, which is
valued on a given Lie algebra $\mathfrak{g}$. In $5D$ it is given by
\begin{equation}
I_{\text{CS}}^{5D }=k\int\left\langle A\left( dA\right) ^{2}+\frac{3}{2}%
A^{3}dA+\frac{3}{5}A^{5}\right\rangle\,,  \label{actm}
\end{equation}
where $\left\langle \cdots\right\rangle $ denotes the invariant tensor for a
given algebra. Remarkably, the invariant tensor for the $\mathfrak{C}_{m}$
algebra (see Eq.~(\ref{invcm})) splits various terms in the action into
different sectors proportional to the arbitrary $\sigma$ constants. As we
are interested in obtaining the maximal PL action, the cosmological constant
and the highest (maximal) order curvature terms both must be in the same
sector. Use of $\mathfrak{C}_{3}=AdS$ and $\mathfrak{C}_{4}=AdS\oplus
Lorentz $ algebras allow us to have these terms in the same sector, however,
the Einstein-Hilbert term is also present there. In fact, these algebras
correspond to the CS Lagrangians having all the gravitational terms (i.e.
purely build from $\omega,e$) proportional just to a single constant. As we
will show now, this is not the case when $\mathfrak{C}_{5}$ is considered,
making it an appropriate candidate to get the EH and the PL terms belonging
to different sectors of the CS action.

To write down the action for the $\mathfrak{C}_{5}$ algebra from the Eq.~(%
\ref{C5}) we start from the connection one-form
\begin{equation}
A=\frac{1}{2}\omega^{ab}J_{ab}+\frac{1}{l}e^{a}P_{a}+\frac{1}{2}k^{ab}Z_{ab}+%
\frac{1}{l}h^{a}R_{a}\,,  \label{1fadsl}
\end{equation}
and the associated curvature two-form
\begin{equation}
F=F^{A}T_{A}=\frac{1}{2}\mathcal{R}^{ab}J_{ab}+\frac{1}{l}\mathcal{T}%
^{a}P_{a}+\frac{1}{2}F^{ab}Z_{ab}+\frac{1}{l}H^{a}R_{a}\,,
\end{equation}
where
\begin{align*}
\mathcal{R}^{ab} & =d\omega^{ab}+\omega_{\text{ }c}^{a}\omega^{cb}+k_{\text{
}c}^{a}k^{cb}+\frac{2}{l^{2}}e^{a}h^{b}\,, \\
& =R^{ab}+k_{\text{}c}^{a}k^{cb}+\frac{2}{l^{2}}e^{a}h^{b}\,, \\
\mathcal{T}^{a} & =de^{a}+\omega_{\text{ }b}^{a}e^{b}+k_{\text{ }%
b}^{a}h^{b}=De^{a}+k_{\text{ }b}^{a}h^{b}\,, \\
& =T^{a}+k_{\text{ }b}^{a}h^{b}\,, \\
F^{ab} & =Dk^{ab}+\frac{1}{l^{2}}h^{a}h^{b}+\frac{1}{l^{2}}e^{a}e^{b}\,, \\
H^{a} & =Dh^{a}+k_{\text{ }b}^{a}e^{b}\,.
\end{align*}

Let us note that the covariant derivative $D=d +\omega$ is only with respect
to the Lorentz part of the connection. See also that $\frac{1}{l^{2}}%
e^{a}e^{b}$ term appears in the $F^{ab}$ curvature instead of being added to
the Lorentz curvature as it usually happens in the AdS gravities. This is a
particular feature of the Maxwell like algebras, which all share the
commutator $\left[ P_{a},P_{b}\right] =Z_{ab}$.

Presence of the $l$ parameter in front of fields $e^{a}$ and $h^{a}$ in Eq.~(%
\ref{1fadsl}) is to make the dimensions right, since our generators are
dimensionless and so must be the gauge connection $A$. As usual in the
general framework of (A)dS-CS theory~\cite{Zan}, this parameter can be
identified with the (A)dS radius $\ell$ introduced in Eq.~(\ref{cosmologicalconstant}), which characterizes the vacuum solutions in the
theory (see also Appendix \ref{App}). Similar identification can be made
when we introduce a gauge connection $A$ valued in other $\mathfrak{C}_{m}$
algebras. Thus, from now on, the scale parameter needed to make the
connection dimensionless always will be give by the AdS radius $\ell$.

Using Theorem VII.2 of Ref.~\cite{Sexp} it is possible to show that the
non-vanishing components of the invariant tensor for the $\mathfrak{C}_{5}$
algebra are given by
\begin{equation}
\begin{tabular}{ll}
$\left\langle J_{ab}J_{cd}P_{e}\right\rangle =\frac{4}{3}\sigma_{1}%
\epsilon_{abcde}\,,$ & $\left\langle J_{ab}J_{cd}Z_{e}\right\rangle =\frac {4%
}{3}\sigma_{3}\epsilon_{abcde}\,,$ \\
$\left\langle J_{ab}Z_{cd}Z_{e}\right\rangle =\frac{4}{3}\sigma_{1}%
\epsilon_{abcde}\,,$ & $\left\langle J_{ab}Z_{cd}P_{e}\right\rangle =\frac {4%
}{3}\sigma_{3}\epsilon_{abcde}\,,$ \\
$\left\langle Z_{ab}Z_{cd}P_{e}\right\rangle =\frac{4}{3}\sigma_{1}%
\epsilon_{abcde}\,,$ & $\left\langle Z_{ab}Z_{cd}Z_{e}\right\rangle =\frac {4%
}{3}\sigma_{3}\epsilon_{abcde}\,,$%
\end{tabular}
\ \ \ \ \ \ \
\end{equation}
where the factor $\frac{4}{3}$ comes from the original invariant tensor of
the AdS algebra and, by the means of the expansion process, is transmitted
to the other components. Note that $\sigma_{1}$ and $\sigma_{3}$ are
arbitrary dimensionless constants allowing to separate the action into two
different sectors. Then considering the gauge connection one-form (\ref{1fadsl}) and the non-vanishing components of the invariant tensor in the
general expression for the $5D$ CS action we find%
\begin{align}
I^{5D}_{\mathfrak{C}_5\text{-CS}}=k\int & \sigma_{1}\left[
\epsilon_{abcde}\left( \frac{1}{\ell}R^{ab}R^{cd}e^{e}+\frac{1}{5\ell^{5}}%
e^{a}e^{b}e^{c}e^{d}e^{e}\right) +\mathcal{\tilde{L}}_{1}\left(
\omega,e,k,h,\right) \right]  \notag \\
+ & \sigma_{3}\left[ \epsilon_{abcde}\left( \frac{2}{3\ell^{3}}%
R^{ab}e^{c}e^{d}e^{e}\right) +\mathcal{\tilde{L}}_{3}\left(
\omega,e,k,h\right) \right] \,,  \label{Acc2}
\end{align}
where $R^{ab}=d\omega^{ab}+\omega_{\text{ }c}^{a}\omega^{cb}$ is usual
Lorentz curvature and $\mathcal{\tilde{L}}_{1}$, $\mathcal{\tilde{L}}_{3}$
are explicitly given by
\begin{align*}
\mathcal{\tilde{L}}_{1} & =\epsilon_{abcde}\left( \frac{2}{\ell}R^{ab}k_{%
\text{ }f}^{c}k^{fd}+\frac{1}{\ell}k_{\text{ }f}^{a}k^{fb}k_{\text{ }%
g}^{c}k^{gd}+\frac{2}{\ell^{5}}h^{a}h^{b}e^{c}e^{d}+\frac{1}{\ell^{5}}%
h^{a}h^{b}h^{c}h^{d}+\frac{1}{\ell}Dk^{ab}Dk^{cd}\right. \\
& \left. +\frac{2}{3\ell^{3}}Dk^{ab}e^{c}e^{d}+\frac{2}{\ell^{3}}%
R^{ab}h^{c}e^{d}+\frac{2}{\ell^{3}}k_{\text{ }f}^{a}k^{fb}h^{c}e^{d}+\frac{2%
}{\ell^{3}}Dk^{ab}h^{c}h^{d}\right) e^{e} \\
& +\epsilon_{abcde}\left( \frac{2}{\ell}Dk^{ab}R^{cd}+\frac{2}{\ell}%
Dk^{ab}k_{\text{ }f}^{c}k^{fd}+\frac{2}{3\ell^{3}}R^{ab}h^{c}h^{d}+\frac {2}{%
3\ell^{3}}k_{\text{ }f}^{a}k^{fb}h^{c}h^{d}\right) h^{e}
\end{align*}
and%
\begin{align*}
\mathcal{\tilde{L}}_{3} & =\epsilon_{abcde}\left( \frac{2}{3\ell^{3}}k_{%
\text{ }f}^{a}k^{fb}e^{c}e^{d}+\frac{1}{\ell^{5}}h^{a}e^{b}e^{c}e^{d}+\frac{2%
}{\ell^{5}}h^{a}h^{b}h^{c}e^{d}+\frac{2}{\ell^{3}}Dk^{ab}h^{c}e^{d}+\frac{2}{%
\ell^{3}}R^{ab}h^{c}h^{d}\right. \\
& \left. +\frac{2}{\ell^{3}}k_{\text{ }f}^{a}k^{fb}h^{c}h^{d}+\frac{2}{\ell }%
Dk^{ab}R^{cd}+\frac{2}{\ell}Dk^{ab}k_{\text{ }f}^{c}k^{fd}\right)
e^{e}+\epsilon_{abcde}\left( \frac{1}{5\ell^{5}}h^{a}h^{b}h^{c}h^{d}\right.
\\
& \left. +\frac{2}{3\ell^{3}}Dk^{ab}h^{c}h^{d}+\frac{1}{\ell}R^{ab}R^{cd}+%
\frac{2}{\ell}R^{ab}k_{\text{ }f}^{c}k^{fd}+\frac{1}{\ell}k_{\text{ }%
f}^{a}k^{fb}k_{\text{ }f}^{c}k^{fd}+\frac{1}{\ell}Dk^{ab}Dk^{cd}\right)
h^{e}\,.
\end{align*}
As we can see, the action (\ref{Acc2}) splits into two pieces. The part
proportional to $\sigma_{1}$ contains the highest order curvature term, the
cosmological constant term and mixed terms with the new extra fields $k^{ab}$ and $h^{a}$ coupled to the spin connection and vielbein. On the
other hand, the piece proportional to $\sigma_{3}$ contains the
Einstein-Hilbert term and a part containing the new extra fields.

After choosing the constant $\sigma_{3}$ to vanish and in a matter-free
configuration $\left( k^{ab}=h^{a}=0\right) $ the action
\begin{equation}
I^{5D}=k\int\sigma_{1}\epsilon_{abcde}\left( \frac{1}{\ell}R^{ab}R^{cd}e^{e}+%
\frac{1}{5\ell^{5}}e^{a}e^{b}e^{c}e^{d}e^{e}\right) \,,  \label{PL1}
\end{equation}
seems to resemble the structure of the maximal Pure Lovelock action (\ref{PLgeneral}) for $p=2$ in $5D$ with the identification $\frac{k\sigma _{1}}{%
\ell}=\kappa$. However, it does not match the sign of the constants in the
desired action. Indeed, using Eq.~(\ref{cts}) we see that these constants
should be equal $\alpha_{2}=\kappa$ and $\alpha_{0}=-\dfrac{1}{5\ell^{4}}%
\kappa$. This relative sign difference makes the PL theory in odd dimensions
and in the torsionless regime to have a unique non degenerate dS and AdS
vacuum (see Ref.~\cite{Dadhich:2015ivt}). This can be directly seen from the
field equations
\begin{equation}
0=\epsilon_{abcde}\left( R^{ab}-\frac{1}{\ell^{2}}e^{a}e^{b}\right) \left(
R^{cd}+\frac{1}{\ell^{2}}e^{c}e^{d}\right) \,.
\end{equation}
One might ask if performing the expansion of the dS algebra might solve this
sign problem but after looking at the full (A)dS Chern-Simons action in five
dimensions,
\begin{equation}
I_{(A)dS\text{-CS}}^{5D}=\kappa\int\epsilon_{abcde}\left( \frac{1}{\ell}%
R^{ab}R^{cd}e^{e}\pm\frac{2}{3\ell^{3}}R^{ab}e^{c}e^{d}e^{e}+\frac{1}{%
5\ell^{5}}e^{a}e^{b}e^{c}e^{d}e^{e}\right) \,,
\end{equation}
it becomes clear that the cosmological term will have exactly the same sign
for AdS and dS symmetries and we will end up with exactly the same action (%
\ref{PL1}).

Naturally, the sign problem in our construction is inherited by the
dynamics. When the previous conditions on the $\sigma$'s and matter fields
are imposed, the field equations of the action (\ref{Acc2}) read,
\begin{align}
0 & =\epsilon_{abcde}\left( R^{ab}R^{cd}+\frac{1}{\ell^{4}}%
e^{a}e^{b}e^{c}e^{d}\right)\,, \\
0 & =\epsilon_{abcde}R^{cd}T^{e}\,, \\
0 & =\epsilon_{abcde}R^{ab}e^{c}e^{d}\,, \\
0 & =\epsilon_{abcde}e^{c}e^{d}T^{e}\,,
\end{align}
which do not correspond to the actual PL dynamics. The system does not even
admit AdS space, $R^{ab}=-\frac{1}{\ell^{2}}e^{a}e^{b}$, as a vacuum
solution. Moreover, the system does not have any maximally symmetric vacuum
solutions. We can only recognize in the last line the torsionless condition.

The same results will be brought by considering the $\mathfrak{C}_{6}$
algebra, which contains more extra fields and it is generated by the set of
generators $\{ J_{ab},P_{a},Z_{ab,(1)},Z_{ab,(2)},R_{a,(1)}\}$ with the
commutation relations given in Ref.~\cite{SS}.

All these facts about signs and dynamics, however, do not mean that the $5D$
case is hopeless. To resolve it we should use a symmetry separating all the
gravity terms into different sectors, instead of 'gluing' some of them
together. Thus, then we could put the whatever sign we want, simply by hand,
since invariant tensors hold arbitrary values. In fact, we do not need to
look very far: we can use for that purpose the $\mathfrak{C}_{7}$ algebra.

This can be treated as a generic method: for particular dimension we should
use some $\mathfrak{C}_{m}$ algebra, with high enough value of the $m$
index, to assure the separation of all purely $(\omega,e)$ terms. That will
bring a much wider framework, not only producing the CS limit to the maximal
PL action but also cases like the EH with $\Lambda$ (see recent Ref.~\cite%
{Gonzalez:2014tta}) and a class of actions similar to the ones discussed in
\cite{Deser:2011zk} having the form $R+LL$, where $R$ is the curvature
scalar and $LL$ is understood as an arbitrary Lanczos-Lovelock term.

Following the definitions of Ref.~\cite{Sexp}, algebra $\mathfrak{C}_{7}$ is
derived from $\mathfrak{so}\left( 4,2\right) $ using $S_{M}^{\left( 5\right)
}$ as the relevant semigroup, whose elements satisfy Eq.~(\ref{ML}). Its
generators $\left\{ J_{ab,\left( i\right) },P_{a,\left( i\right) }\right\}$,
with index $i=0,1,2$ numerating different types, will be related to the $%
\mathfrak{so}\left( 4,2\right) $ ones by%
\begin{align*}
J_{ab,\left( i\right) } & =\lambda_{2i}\tilde{J}_{ab}\,, \\
P_{a,\left( i\right) } & =\lambda_{2i+1}\tilde{P}_{a}\,,
\end{align*}
and will satisfy Eq.~(\ref{Cm}). Let now $A$ be the connection one-form for $%
\mathfrak{C}_{7}$,%
\begin{equation}
A=\frac{1}{2}\omega^{ab,\left( i\right) }J_{ab,\left( i\right) }+\frac {1}{%
\ell}e^{a,\left( i\right) }P_{a,\left( i\right) }\,,  \label{COF}
\end{equation}
where the spin connection and vielbein being inside are defined as $%
\omega^{ab}=\omega^{ab,\left( 0\right) }$ and $e^{a}=e^{a,\left( 0\right) }$%
, respectively. The extra fields will be denoted as $k^{ab,(i)}=\omega^{ab,(i)}$ and $h^{a,(i)}=e^{a,(i)}$, with $i \neq 0$.

Using Theorem VII.2 of Ref.~\cite{Sexp}, it is possible to show that the
non-vanishing components of the invariant tensor are given by%
\begin{equation}
\left\langle J_{ab,\left( q\right) }J_{cd,\left( r\right) }P_{e,\left(
s\right) }\right\rangle =\frac{4}{3}\sigma_{2u+1}\delta_{j\left(
q,r,s\right) }^{u}\epsilon_{abcde}\,,  \label{ITC7}
\end{equation}
where the $\sigma$'s are arbitrary constants and $j\left( q,r,s\right) $ for
$m=7$ satisfies Eq.~(\ref{rule}). We express the corresponding $5D$ CS
action as
\begin{equation}
I_{\mathfrak{C}_7\text{-CS}}^{5D}=k\int\sum_{i=0}^{2}\sigma_{2i+1}\mathcal{L}%
_{2i+1}\,,
\end{equation}
where
\begin{align}
\mathcal{L}_{2i+1} & =\epsilon_{abcde}\left( \frac{\delta_{j\left(
q,r,s\right) }^{i}}{\ell}R^{ab,\left( q\right) }R^{cd,\left( r\right)
}e^{\left( s\right) }+\frac{2\delta_{j\left( q,r,s,u,1\right) }^{i}}{%
3\ell^{3}}R^{ab,\left( q\right) }e^{c,\left( r\right) }e^{d,\left( s\right)
}e^{e,\left( u\right) }\right.  \notag \\
& \left. +\frac{\delta_{j\left( q,r,s,u,v,2\right) }^{i}}{5\ell^{5}}%
e^{a,\left( q\right) }e^{b,\left( r\right) }e^{c,\left( s\right)
}e^{d,\left( u\right) }e^{e,\left( v\right) }\right) \,,  \label{CS5C7_term}
\end{align}
and $j( \ldots) $ is satisfying Eq.~(\ref{rule}). The explicit form of the
action expression can be found in Appendix \ref{App2}. After writing the
purely gravitational terms $( \omega,e) $ separately from those containing
extra fields we obtain
\begin{align}
I_{\mathfrak{C}_7\text{-CS}}^{5D}=k\int & \sigma_{1}\left[
\epsilon_{abcde}\left( \frac{1}{\ell }R^{ab}R^{cd}e^{e}\right) +\mathcal{%
\tilde{L}}_{1}\left( \omega^{\left( i\right) },e^{\left( j\right) }\right) %
\right] +\sigma_{3}\left[ \epsilon_{abcde}\left( \frac{2}{3\ell^{3}}%
R^{ab}e^{c}e^{d}e^{e}\right) +\mathcal{\tilde{L}}_{3}\left( \omega^{\left(
i\right) },e^{\left( j\right) }\right) \right]  \notag \\
& +\sigma_{5}\left[ \epsilon_{abcde}\left( \frac{1}{5\ell^{5}}%
e^{a}e^{b}e^{c}e^{d}e^{e}\right) +\mathcal{\tilde{L}}_{5}\left(
\omega^{\left( i\right) },e^{\left( j\right) }\right) \right] \,.
\label{CS5C7}
\end{align}

We can see that $\mathfrak{C}_{7}$ allows us to have each term in a
different sector. When the $\sigma_{3}$ constant vanishes and $%
\sigma_{5}=-\sigma_{1}$, the matter-free configuration limit $\left(
k^{ab,(i)}=h^{a,(j)}=0\right)$ leads to the maximal $p=2$ Pure Lovelock action
\begin{equation}
I_{\text{CS}\to \text{PL}}^{5D}=k\int\sigma_{1}\epsilon_{abcde}\left( \frac{1%
}{\ell}R^{ab}R^{cd}e^{e}-\frac{1}{5\ell^{5}}e^{a}e^{b}e^{c}e^{d}e^{e}\right)
\,.  \label{truePL}
\end{equation}

An additional non-trivial choice of the constants to consider is $%
\sigma_{1}=0$ with $\sigma_{3}=\sigma_{5}$. This particular choice leads to
the $p=1$ Pure Lovelock corresponding to EH gravity with cosmological
constant. Thus, we have shown that the PL actions of different orders can be
related to the CS action for the $\mathfrak{C}_{7}$ algebra. We remind that
these choices of the $\sigma$'s are allowed since they are all arbitrary.

Unfortunately, although matching form of the actions, this still does not
lead to the PL dynamics alone. When the constant $\sigma _{3}$ vanishes, a
matter-free configuration is considered and $\sigma _{5}=-\sigma _{1}$, the
field equations read respectively,
\begin{align}
0& =\epsilon _{abcde}\left( R^{ab}R^{cd}-\frac{1}{\ell ^{4}}%
e^{a}e^{b}e^{c}e^{d}\right) \delta e^{e},  \notag \\
0& =\epsilon _{abcde}\left( \frac{2}{\ell ^{2}}R^{ab}e^{c}e^{d}-\frac{1}{%
\ell ^{4}}e^{a}e^{b}e^{c}e^{d}\right) \delta h^{e,\left( 1\right) },  \notag
\\
0& =\epsilon _{abcde}\left( R^{ab}R^{cd}-\frac{2}{\ell ^{2}}%
R^{ab}e^{c}e^{d}\right) \delta h^{e,\left( 2\right) }, \label{field_eq_5D_C7}
\end{align}
and
\begin{align}
0& =\epsilon _{abcde}\left( R^{cd}T^{e}\right) \delta \omega ^{ab},  \notag
\\
0& =\epsilon _{abcde}\left( \frac{2}{\ell ^{2}}e^{c}e^{d}T^{e}\right) \delta
k^{ab,\left( 1\right) },  \notag \\
0& =\epsilon _{abcde}\left( R^{cd}T^{e}-\frac{2}{\ell ^{2}}%
e^{c}e^{d}T^{e}\right) \delta k^{ab,\left( 2\right) }\,.
\label{field_eq_5D_C7b}
\end{align}%
As it was pointed out in the Introduction, if we use the first order
formalism to describe PL gravity, then the vanishing of torsion must be
imposed by hand (or by introducing suitable Lagrange multiplier \cite%
{Dadhich:2015ivt}). Here instead, this condition is coming from the
variation of the extra $\delta k^{ab,(1)}$ field.

When the matter fields are switched off and torsionless condition is assumed
the solution still has to simultaneously satisfy the PL and two other
unusual equations. This problem has already appeared in Ref.~\cite{Zan}, but
there one finds superposition of vanishing of the $RR$ and $Reee$ terms.
Such restriction of the geometry rejected possibility of the spherical
solutions but at least it was fulfilled by the pp-waves.

One can clearly see that proper handling of the full problem requires a more
subtle treatment. Nevertheless, if we consider at the level of the action (\ref{CS5C7}) some special extra
fields identification among extra fields:
\begin{equation}
h^{a,\left( 1\right) }=h^{a,\left( 2\right) }=h^{a}\text{ \ \ and\ \ \ }%
k^{ab,\left( 1\right) }=k^{ab,\left( 2\right) }=k^{ab}\,,  \label{ident}
\end{equation}%
and later impose $\sigma _{3}=0$, $\sigma _{5}=-\sigma
_{1} $, then we will be able to find the following field equations in a matter-free configuration
limit:%
\begin{align}
0& =\epsilon _{abcde}\left( R^{ab}R^{cd}-\frac{1}{\ell ^{4}}%
e^{a}e^{b}e^{c}e^{d}\right) \delta e^{e},  \notag \\
0& =\epsilon _{abcde}\left( R^{ab}R^{cd}-\frac{1}{\ell ^{4}}%
e^{a}e^{b}e^{c}e^{d}\right) \delta h^{e},  \notag \\
0& =\epsilon _{abcde}\left( R^{cd}T^{e}\right) \delta \omega ^{ab},  \notag
\\
0& =\epsilon _{abcde}\left( R^{cd}T^{e}\right) \delta k^{ab}\,.
\end{align}%
Thus, we obtain the appropriate PL dynamics described in Refs.~\cite%
{Cai:2006pq, Dadhich:2012ma, Dadhich:2012eg}. Note that although we are using the same $\sigma$'s constants and eventually we enforce the extra matter free configuration limit, the PL dynamics is recovered only from the Lagrangian $\mathcal{L}(\omega, e,h,k)$ and not from the $\mathcal{L}(\omega, e, k^{(1)}, h^{(1)}, k^{(2)}, h^{(2)})$. Interestingly, proposed type of identification on the matter field from (\ref{ident}) can be applied not only in a 5D CS action but also in higher dimensional cases.

One could try to interpret $\,h^{a}$ as the vielbein, nevertheless it is straightforward to see that the action (%
\ref{CS5C7}) reproduces the PL action only when $e^{a}$ is identified as the
true vielbein. Moreover, the matter extra fields ($h^{a},k^{ab}$)
could allow to introduce a generalized cosmological term to the CS gravity
action in an analogous way to the one introduced in the four-dimensional case \cite{SS, AKL}.

For the completeness, we will now supply the
generalization of our construction to higher dimensions.

\section{The higher-dimensional Pure Lovelock and Chern-Simons gravity
actions}

In this section we present the general setup to obtain the $( 2n+1)$%
-dimensional PL action for any value of $p$ from CS gravity theory using the
Maxwell type $\mathfrak{C}_{m}$ family. In particular, we show that the $%
\mathfrak{C}_{7}$ algebra for $7D$ allows us to recover the maximal $p=3$ PL
action, where the sign problem from the five-dimensional case does not occur.

Let us remind the generic form of the PL constants%
\begin{equation}
\alpha_{p}=\frac{1}{(D-2p)!}\,\kappa\text{\qquad and \qquad}\alpha _{0}=-%
\dfrac{\left( \mp1\right) ^{p}}{D\left( D-2p-1\right) !\ell^{2p}}\,\kappa\,,
\end{equation}
with AdS ($-$ sign) and the dS ($+$ sign), and compare them with the generic
(A)dS-Chern-Simons ones \cite{Zanelli:2005sa,TZ}
\begin{equation}
\tilde{\beta}_{p}=\frac{\left( -sgn\left( \Lambda\right) \right) ^{p}}{%
(D-2p)\ell^{D-2p}}\binom{\frac{D-1}{2}}{p}\,k\text{\qquad and \qquad}\tilde{%
\beta}_{0}=\frac{1}{D\ell^{D}}\,k\,.
\end{equation}
Notice that for the PL constants $p$ is fixed, while for the CS constants
index $p$ is running from $0$ to the maximum power of the curvature $N=[%
\frac {D-1}{2}]$. The construction of the $D=2n+1$ dimensional $\mathfrak{C}%
_{m}$-Chern-Simons have to take into account also the arbitrary constants $%
\sigma$%
\begin{equation*}
\beta_{p}=\frac{\left( -sgn\left( \Lambda\right) \right) ^{p}}{(D-2p)\ell
^{D-2p}}\binom{\frac{D-1}{2}}{p}k\ \sigma_{2i+1}\delta_{j\left(
i_{1},\ldots,i_{D-p},\frac{D-1}{2}-p\right) }^{i}\text{ \ and \ }\beta_{0}=%
\frac {1}{D\ell^{D}}k\,\sigma_{2i+1}\delta_{j\left( i_{1},\ldots,i_{D},\frac{%
D-1}{2}\right) \ }^{i},
\end{equation*}
where algebraic $m$ dependence appears explicitly in the definition of
\begin{equation*}
j\left( i_{1},i_{2},\ldots,i_{n+1}\right) =\left( i_{1}+i_{2}+\cdots
+i_{n+1}\right) \operatorname{mod}\left( \frac{m-1}{2}\right) \,,
\end{equation*}
and is responsible for shifting terms into different sectors of the
invariant tensor. Then restricting only to the purely gravitational terms we
obtain%
\begin{align*}
PL^{(A)dS} & :\ \ \frac{\alpha_{0}}{\alpha_{p}}=-\left( \mp1\right) ^{p}%
\dfrac{(D-2p)}{D}\frac{1}{\ell^{2p}}\,, \\
CS^{\mathfrak{C}_{m}} & :\ \ \frac{\beta_{0}}{\beta_{p}}=\left( -sgn\left(
\Lambda\right) \right) ^{p}\frac{(D-2p)}{D}\frac{1}{\ell^{2p}}\frac {1}{%
\binom{\frac{D-1}{2}}{p}}\frac {\delta_{n\operatorname{mod}\frac{m-1}{2}}^{i}}{%
\delta_{\left( n-p\right) \operatorname{mod}\frac{m-1}{2}}^{j}}\frac{\sigma_{2i+1}}{%
\sigma_{2j+1}}\,,
\end{align*}
from where we see that%
\begin{equation}
\left( -sgn\left( \Lambda\right) \right) _{CS}^{p}\frac{1}{\binom{n}{p}}%
\frac{\sigma_{2(n\operatorname{mod}\frac{m-1}{2})+1}}{\sigma_{2(\left( n-p\right)
\operatorname{mod}\frac{m-1}{2})+1}}=-\left( \mp1\right) _{PL}^{p}\ .
\end{equation}
For the same $\sigma$'s, whose cancellation forces $\binom{n}{p}=1$, we are
restricted only to the maximal $p=N$ PL. Then we see that $p=even$ leads to
a sign contradiction, as we have found explicitly in the previous section.
This problem does not appear when maximal order happens to be $p=odd$, which
is possible only for $D=3, 7, 11, \ldots, 4k-1$. For other situations it is
necessary to absorb the problematic sign and the numerical factors into
definition of one of the $\sigma$'s.

Indeed, in Section 2 we have shown that the five-dimensional CS action
constructed using $\mathfrak{C}_{5}$ could not lead to the maximal $p=2$ PL
action with the proper sign but could only produce the right $p=1$ PL
action, which is just the Einstein-Hilbert action with $\Lambda$. Finally,
we managed to show that $\mathfrak{C}_{7}$ algebra in that context allowed
us to accommodate minus sign by the means of setting $\sigma_{5}=-\sigma_{1}$%
, which effectively provides good maximal $p=2$ PL action.

On the other hand, the right sign for the maximal $p=3$ PL in $7D$ will be
obtained in a straightforward way, simply by using $\mathfrak{C}_{7}$.
Additionally, by using $\mathfrak{C}_{9}$ algebra, we can derive other
seven-dimensional PL actions, $p=2$ and $p=1$. Naturally $p=3$ can be also
establish within that bigger algebra. However, taking into account growing
complexity with the transition to the higher value of $m$ (see for example
Appendix \ref{App2}), we will be always targeting in the minimal setup
leading to the right result.\newline

\noindent Then the full picture in odd dimensions becomes clear and it can
be summarized in the following way:

\begin{itemize}
\item for $D=4k-1$, the smallest representative in the $\mathfrak{C}_{m}$
family that allows to obtain maximal $p=N$ PL action is given by $\mathfrak{C%
}_{D}$, whereas for any other order of PL $p=1,\ldots,(N-1)$ one needs to
use $\mathfrak{C}_{D+2}$,

\item for $D\neq4k-1$, the smallest representative in the $\mathfrak{C}_{m}$
family that allows to obtain arbitrary order of PL action for $p=1,\ldots,N$
is given by $\mathfrak{C}_{D+2}$.
\end{itemize}

Obviously, all these PL actions could be recovered as a limit using even
bigger $\mathfrak{C}_{m}$ algebras, but then one needs to introduce more extra fields and conditions on the $\sigma$'s.

Although the PL actions can be obtained from the CS theory, their right dynamical limit requires appropriate identifications of the extra fields at the level of the action in order to reproduce the PL dynamics:
\begin{align}
h^{a,(1)}&=h^{a,(2i+1)}=(\pm 1)^{p}h^{a,(2i)} \, , \notag \\
k^{ab,(1)}&=k^{ab,(2i+1)}=(\pm 1)^{p}k^{ab,(2i)}\,,
\end{align}
for all $i=1,2,\dots,\frac{m-3}{2}$.

One should notice that even if we consider $dS$ case as a starting point it would lead to the same sign problems, as was mentioned in five-dimensional case.
\subsection{Chern-Simons gravity and Pure Lovelock action in $D=7$}

The generic form of a CS action in seven dimensions \cite{Cham1, Cham2} is
given by
\begin{equation}
I_{\text{CS}}^{7D}=k\int\left\langle A\left( dA\right) ^{3}+\frac{8}{5}%
A^{3}\left( dA\right) ^{2}+\frac{4}{5}A\left( dA\right) A^{2}\left(
dA\right) +2A^{5}dA+\frac{4}{7}A^{7}\right\rangle \,.
\end{equation}

Let us first consider the $\mathfrak{C}_{7}=\left\{ J_{ab,\left( i\right)
},P_{a,\left( i\right) }\right\}$ algebra, whose generators satisfy Eq.~(\ref{Cm}), with $i=0,1,2$. The associated connection one-form is defined as
\begin{equation}
A=\frac{1}{2}\omega^{ab,\left( i\right) }J_{ab,\left( i\right) }+\frac {1}{%
\ell}e^{a,\left( i\right) }P_{a,\left( i\right) }\,.  \label{1f7}
\end{equation}
From Theorem VII.2 of Ref.~\cite{Sexp} and according to Eq.~(\ref{invcm}),
it is possible to show that the non-vanishing components of the invariant
tensor of order 4 for $\mathfrak{C}_{7}$ are given by%
\begin{equation}
\left\langle J_{ab,\left( q\right) }J_{cd,\left( r\right) }J_{ef,\left(
s\right) }P_{g,\left( u\right) }\right\rangle =2\sigma_{2v+1}\delta_{j\left(
q,r,s,u\right) }^{v}\epsilon_{abcdefg}\,,  \label{invt7}
\end{equation}
where the $\sigma$'s are arbitrary constants and $j\left( q,r,s,u\right) $
satisfies Eq.~(\ref{rule}). Then using the invariant tensor for $\mathfrak{C}%
_{7}$ (\ref{ITC7}) and introducing the gauge connection one-form (\ref{1f7})
in the general expression for the $7D$ CS action gives
\begin{equation}
I_{\mathfrak{C}_7\text{-CS}}^{7D}=k\int\sum_{i=0}^{2}\sigma_{2i+1}\mathcal{L}%
_{2i+1}\,,
\end{equation}
where
\begin{align*}
\mathcal{L}_{2i+1} & =\epsilon_{abcdefg}\left( \frac{\delta_{j\left(
q,r,s,u\right) }^{i}}{\ell}R^{ab,\left( q\right) }R^{cd,\left( r\right)
}R^{ef,\left( s\right) }e^{g,\left( u\right) }+\frac{\delta_{j\left(
q,r,s,u,v,1\right) }^{i}}{3\ell^{3}}R^{ab,\left( q\right) }R^{cd,\left(
r\right) }e^{e,\left( s\right) }e^{f,\left( u\right) }e^{g,\left( v\right)
}\right.  \notag \\
& +\frac{3\delta_{j\left( q,r,s,u,v,w,2\right) }^{i}}{5\ell^{5}}R^{ab,\left(
q\right) }e^{a,\left( r\right) }e^{b,\left( s\right) }e^{c,\left( u\right)
}e^{d,\left( v\right) }e^{e,\left( w\right) }\,  \notag \\
& \left. +\frac{\delta_{j\left( q,r,s,u,v,w,o,3\right) }^{i}}{7\ell^{7}}%
e^{a,\left( q\right) }e^{b,\left( r\right) }e^{c,\left( s\right)
}e^{d,\left( u\right) }e^{e,\left( v\right) }e^{f,\left( w\right)
}e^{g,\left( o\right) }\right) \,.
\end{align*}
Separating the purely gravitational terms from those containing extra
fields, the action can be written as
\begin{align}
I^{7D}_{\mathfrak{C}_7\text{-CS}}=k & \int\sigma_{1}\left[
\epsilon_{abcdefg}\left( \frac{1}{\ell}R^{ab}R^{cd}R^{ef}e^{g}+\frac{1}{%
7\ell^{7}}e^{a}e^{b}e^{c}e^{d}e^{e}e^{f}e^{g}\right) +\mathcal{\tilde{L}}%
_{1}\left( \omega^{\left( i\right) },e^{\left( j\right) }\right) \right]
\notag \\
& +\sigma_{3}\left[ \epsilon_{abcdefg}\left( \frac{1}{\ell^{3}}%
R^{ab}R^{cd}e^{e}e^{f}e^{g}\right) +\mathcal{\tilde{L}}_{3}\left(
\omega^{\left( i\right) },e^{\left( j\right) }\right) \right]  \notag \\
& +\sigma_{5}\left[ \epsilon_{abcdefg}\left( \frac{3}{5\ell^{5}}%
R^{ab}e^{c}e^{d}e^{e}e^{f}e^{g}\right) +\mathcal{\tilde{L}}_{5}\left(
\omega^{\left( i\right) },e^{\left( j\right) }\right) \right] \,.
\label{CS7C7}
\end{align}
Let us notice that the action (\ref{CS7C7}) splits into three pieces. The
part proportional to $\sigma_{1}$ contains the maximal PL Lagrangian and
mixed terms containing extra fields $k^{ab,(i)
}=\omega^{ab,(i)}$ and $h^{a,(i)}=e^{a,(i)}$, with $i \neq 0$. Unlike the five-dimensional case, when
the constants $\sigma_{3}\,$, $\sigma_{5}$ vanish a matter-free
configuration $\left(k^{ab,(i)}=h^{a,(i)}=0\right)$ leads to the action corresponding to the maximal $%
p=3$ PL:
\begin{equation}
I^{7D}_{\text{CS}\to\text{PL}}=k\int\sigma_{1}\epsilon_{abcdefg}\left( \frac{%
1}{\ell}R^{ab}R^{cd}R^{ef}e^{g}+\frac{1}{7\ell^{7}}%
e^{a}e^{b}e^{c}e^{d}e^{e}e^{f}e^{g}\right) \,.
\end{equation}

Following the same procedure for the bigger $\mathfrak{C}_{9}$ algebra we
extend the form of the connection (\ref{1f7}) to the range of $i=0,1,2,3$,
which also makes
\begin{equation*}
I_{\mathfrak{C}_9\text{-CS}}^{7D}=k \int\sum_{i=0}^{3}\sigma_{2i+1}\mathcal{L%
}_{2i+1}\,.
\end{equation*}
That gives the $\mathfrak{C}_{9}$-CS action, which now splits into four
terms
\begin{align}
I_{\mathfrak{C}_9\text{-CS}}^{7D}= & k \int\sigma_{1}\left[
\epsilon_{abcdefg}\left( \frac{1}{\ell}R^{ab}R^{cd}R^{ef}e^{g}\right) +%
\mathcal{\tilde{L}}_{1}\left( \omega^{\left( i\right) },e^{\left( j\right)
}\right) \right]  \notag \\
& +\sigma_{3}\left[ \epsilon_{abcdefg}\left( \frac{1}{\ell^{3}}%
R^{ab}R^{cd}e^{e}e^{f}e^{g}\right) +\mathcal{\tilde{L}}_{3}\left(
\omega^{\left( i\right) },e^{\left( j\right) }\right) \right]  \notag \\
& +\sigma_{5}\left[ \epsilon_{abcdefg}\left( \frac{3}{5\ell^{5}}%
R^{ab}e^{c}e^{d}e^{e}e^{f}e^{g}\right) +\mathcal{\tilde{L}}_{5}\left(
\omega^{\left( i\right) },e^{\left( j\right) }\right) \right]  \notag \\
& +\sigma_{7}\left[ \epsilon_{abcdefg}\left( \frac{1}{7\ell^{7}}%
e^{a}e^{b}e^{c}e^{d}e^{e}e^{f}e^{g}\right) +\mathcal{\tilde{L}}_{7}\left(
\omega^{\left( i\right) },e^{\left( j\right) }\right) \right] \,.
\label{CS7C9}
\end{align}
Interestingly, each term of the original seven-dimensional AdS-CS action
appears in a different sector of the $\mathfrak{C}_{9}$-CS action. This
feature allows us to reproduce any $p$-order of the PL theory, as it
happened for $\mathfrak{C}_{7}$ algebra in five-dimensional case. Each order
can be derived in a matter-free configuration, after imposing the following
conditions:
\begin{align}
p=1&: & \sigma_{1}=\sigma_{3}=0\,, & & \sigma_{5}=\sigma_{7}\,,  \label{p1PL}
\\
p=2&: & \sigma_{1}=\sigma_{5}=0\,, & & -\sigma_{3}=\sigma_{7}\,,
\label{p2PL} \\
p=3&: & \sigma_{3}=\sigma_{5}=0\,, & & \sigma_{1}=\sigma_{7}\,.  \label{p3PL}
\end{align}
The first case reproduces the EH with cosmological constant, the second
represents first nontrivial $7D$ PL, while third one gives the maximal PL.
Note that last result was already achieved by the smaller $\mathfrak{C}_{7}$
algebra, as the spacetime dimension is a particular case of $D=4k-1$.

When we turn to the dynamical limit of (\ref{p1PL})-(\ref{p3PL}) we face
very complicated superposition of the field equations. Only minimal and
maximal cases, $p=1$ and $p=3$ automatically assure the torsionless
condition. For the other intermediate case, $p=2$, vanishing of torsion is
not coming from field equations. On the other hand, the following
identification in the action (\ref{CS7C9}):%
\begin{eqnarray}
h^{a,\left( 1\right) } &=&h^{a,\left( 3\right) }=\left( \pm 1\right)
^{p}h^{a,\left( 2\right) }\,, \notag \\
k^{ab,\left( 1\right) } &=&k^{ab,\left( 3\right) }=\left( \pm 1\right)
^{p}k^{ab,\left( 2\right) }\,,
\end{eqnarray}%
reproduces the desired PL dynamics for any $p$-order after considering the
appropriate conditions on the $\sigma $'s and in a matter-free configuration
limit.

One might try to look for a modification of the $\mathfrak{C}_{m}$ algebra
that, besides giving the right limit for the action, it leads also to the
right dynamical limit without any identification on the extra fields. However, we will leave that task for a future work.

It is important to point out that other non-trivial conditions on the $%
\sigma $'s can lead to other interesting gravity theories. In fact, after
choosing $\sigma_{5}=\sigma_{1}$ or $\sigma_{5}=\sigma_{3}$ and killing all
other $\sigma$ constants, a matter-free configuration will reproduce the
class of actions discussed in Ref.~\cite{Deser:2011zk} having the form $EH+LL
$, where $LL$ corresponds to an arbitrary Lanczos-Lovelock term.

\subsection{Chern-Simons gravity and Pure Lovelock action in $D=2n+1$}

To assure the arbitrary $p$-order PL action resulting from $D=2n+1$ CS
gravity action we require the use of $\mathfrak{C}_{D+2}$ algebra. As
previously, we start with the connection one-form
\begin{equation}
A=\frac{1}{2}\omega^{ab,\left( i\right) }J_{ab,\left( i\right) }+\frac {1}{%
\ell}e^{a,\left( i\right) }P_{a,\left( i\right) }\,,
\end{equation}
with $i=0,\ldots,n$ and the non-vanishing components of the invariant tensor
given by Eq.~(\ref{invcm}). It is possible to show that the $(2n+1)$%
-dimensional CS action for the $\mathfrak{C}_{D+2}$ algebra is then given by%
\begin{align}
I^{2n+1}_{\mathfrak{C}_{D+2}\text{-CS}} & =k\int\epsilon_{a_{1}a_{2}\ldots
a_{2n+1}}\sum_{p=0}^{n}\frac {\ell^{2\left( p-n\right) -1}}{2\left(
n-p\right) +1}\binom{n}{p}\sigma_{2i+1}\delta_{j\left(
i_{1},\ldots,i_{2n+1-p},n-p\right) }^{i}R^{a_{1}a_{2},\left( i_{1}\right)
}\cdots R^{a_{2p-1}a_{2p},\left( i_{p}\right) }  \notag \\
& \times e^{a_{2p+1},\left( i_{p+1}\right) }\cdots e^{a_{2n+1},\left(
i_{2n+1-p}\right) }\,,  \label{CSG}
\end{align}
where we identify $\omega^{ab}=\omega^{ab,\left( 0\right) }$, $%
e^{a}=e^{a,\left( 0\right) }$ and function $j(\ldots)$ in the Kronecker
delta satisfies Eq.~(\ref{rule}). The action (\ref{CSG}) is separated into $%
n+1$ pieces proportional to the different $\sigma$'s and under specific
conditions ($\sigma_{2n+1-2p}=(\pm 1)^{p+1}\sigma_{2n+1}$, vanishing all other $\sigma$
constants and of course applying the free-matter configuration limit) gives $%
p$-order PL action
\begin{align}
I^{2n+1}_{\text{PL}} & =k\int\sigma_{2n+1-2p}\ \epsilon_{a_{1}a_{2}\ldots
a_{2n+1}}\left( \frac{\ell^{2\left( p-n\right) -1}}{2\left( n-p\right) +1}%
\binom{n}{p}R^{a_{1}a_{2}}\cdots R^{a_{2p-1}a_{2p}}e^{a_{2p+1}}\cdots
e^{a_{2n+1}}\right.  \notag \\
& \left. +\frac{\ell^{-2n-1}}{\left( 2n+1\right) }e^{a_{1}}e^{a_{2}}\cdots
e^{a_{2n+1}}\right) \,.
\end{align}
Naturally, for $D=4k-1$ to derive the maximal PL action from CS gravity
theory it is enough to use the $\mathfrak{C}_{D}$ algebra. The only
difference will appear in the range of the indices \mbox{$i=0,\ldots,n-1$}
enumerating various types of generators and, consequently, in the modulo
function inside the definition of $j(\ldots)$ present in the Eq.~(\ref{CSG}).

Interestingly, we observe that the extra field content $k^{ab,(i)}$ for
minimal $p=1$ (EH) and maximal $p=N$ PL always leads to the explicit
torsionless condition, whereas for other situations the torsion remains
still involved with the rest of the field equations.

Arbitrariness of the $\sigma$'s allows us to construct a wide class of
actions, where we can couple arbitrary $LL$ terms together. One of the
notable examples could be the Einstein-Hilbert term equipped by the
arbitrary $p$-order $LL$ term \cite{Deser:2011zk} (coming from setting $%
\sigma_{2n+1-2p}=\sigma_{2n-1}$ for $p\neq0\,$ and vanishing all others $%
\sigma$'s)
\begin{align}
I_{\text{EH+LL}}^{2n+1} & =k\int\sigma_{2n+1-2p}\ \epsilon_{a_{1}a_{2}\ldots
a_{2n+1}}\left( \frac{n\ell^{1-2n}}{2n-1}R^{a_{1}a_{2}}e^{a_{3}} \cdots
e^{a_{2n+1}}\right.  \notag \\
& \left. +\frac{\ell^{2\left( p-n\right) -1}}{2\left( n-p\right) +1}\binom{n%
}{p} R^{a_{1}a_{2}} \cdots R^{a_{2p-1}a_{2p}}e^{a_{2p+1}}\cdots
e^{a_{2n+1}}\right) \,.
\end{align}

\section{Conclusion}

We have managed to explore applications of the $\mathfrak{C}_{m}$ algebras
in a CS gravity theory to incorporate the PL actions as a special limit. It
was achieved by the fact that the $\mathfrak{C}_{m}$ algebra is shifting
various powers of the Riemann tensor into different sectors of the invariant
tensor. Although it allowed us to obtain the proper $p$-order PL action in
arbitrary spacetime dimension (by particular choice of the algebra and by
manipulating the constants), we faced a problem in the dynamical limit
caused by additional field equations, similarly as happens in Ref.~\cite{Zan}
in the context of CS and GR. For the maximal and minimal case we see that the extra fields give the explicit torsionless condition, which can be an interesting feature. Finally to overcome the non-trivial superposition of the field equations overconstraining standard solutions we have proposed a particular identification of the matter fields at the level of the action allowing to reproduce the correct PL dynamics for any value of $p$.

Altogether the general treatment involves for arbitrary odd $D$-dimensional
spacetime the use of $\mathfrak{C}_{D+2}$ as it allows us to fully
manipulate the gravitational terms through imposing suitable conditions on
the $\sigma$ constants. In this way we obtained not only the maximal but
also arbitrary order of the PL action along with the gravity theories
similar to those discussed in Ref.~\cite{Deser:2011zk}. Some cases can be
achieved with less effort. For example in $D=4k-1$-dimensions it is
sufficient to use the $\mathfrak{C}_{D}$ algebra to get the maximal PL.

The possibility of finding a suitable symmetry leading also to the
right dynamical limit with less conditions on the extra fields still remains
as an open problem. This will be approached in a future work with techniques
developed in Refs.~\cite{CarocaNelson,AMNT}. Finally, the mechanism presented here could be useful to derive the supersymmetric version of the PL theory.

\section{Acknowledgment}

This work was supported by the Chilean FONDECYT Projects No. 3140267 (RD)
and 3130445 (NM), and also funded by the Newton-Picarte CONICYT Grant No.
DPI20140053 (PKC and EKR). PKC and EKR wish to thank A. Anabal\'{o}n for his
kind hospitality at Departamento de Ciencias of Universidad Adolfo Iba\~{n}%
ez. RD and NM would also like to thank J. Zanelli for valuable discussion and comments.

{\small \appendix}

\section{Appendix\label{App}}

Here we present the explicit relation between the constant $k$ appearing in
the $5D$ CS action and the invariant tensor constants.

The Pure Lovelock theory is described by the action%
\begin{align}
I^{5D}_{\text{PL}} & =\kappa\int\epsilon_{abcde}\left( \frac{1}{\ell}%
R^{ab}R^{cd}e^{e}-\frac{1}{5\ell^{5}}e^{a}e^{b}e^{c}e^{d}e^{e}\right) \,,
\label{PL1a}
\end{align}
which is equivalent to the tensorial action%
\begin{align}
I^{5D}_{\text{PL}} & =-\kappa\int\sqrt{-g}\left( \,\frac{1}{4}%
R_{\mu\nu}^{\rho\sigma}R_{\alpha\beta}^{\gamma\lambda}\delta_{\rho\sigma%
\gamma\lambda }^{\mu\nu\alpha\beta}-2\Lambda\right) d^{5}x\ .
\end{align}
When dealing with solutions on this theory \cite%
{Cai:2006pq,Dadhich:2012eg,Dadhich:2012ma}, in both formalisms, the
vanishing of torsion is usually imposed by hand, either by $T^{a}=0$ or $%
T^{\alpha}_{\mu\nu}=0$.

We, however, are interested in finding this action as a special limit of a
CS action constructed for a specific symmetry,%
\begin{equation*}
I _{\text{CS}}^{5D} =k\int\left\langle A\left( dA\right) ^{2}+\frac{3}{2}%
A^{3}dA+\frac{3}{5}A^{5}\right\rangle \,,
\end{equation*}
where $A$ is the gauge connection valued on the corresponding algebra and$\
k $ (different, in principle, from $\kappa$) is a constant that makes this
functional of $A$ have dimension of the action. We notice that all the
factors are dimensionless, as the connection is constructed in a way it is
dimensionless too. When the algebra is chosen to be $\mathfrak{C}_{7}$, then
the connection is given by
\begin{equation*}
A=\frac{1}{2}\omega^{ab}J_{ab}+\frac{1}{l}e^{a}P_{a}+\frac{1}{2}%
k^{ab,(1)}Z_{ab,(1)}+\frac{1}{l}h^{a,(1)}R_{a,(1)}+\frac{1}{2}%
k^{ab,(2)}Z_{ab,(2)}+\frac{1}{l}h^{a,(2)}R_{a,(2)}\,,
\end{equation*}
where the length parameter $l$ assures that $A$ is indeed dimensionless
(remember that vielbein has dimension of length, and then extra fields like $%
h^{a,(i)}$ acquire the same dimension).

To analyze the relation between $\mathfrak{C}_{7}$-CS and PL actions, we
need to identify $l$ with the AdS radius $\ell$ characterizing a vacuum
solution of the theory, as it is usually done in (A)dS-CS theory~\cite{Zan}.
In Section 3 we showed that effectively this leads to
\begin{align}
I^{5D}_{\text{CS}\to \text{PL}} =k\sigma_{1}\int\epsilon_{abcde}\left( \frac{%
1}{\ell}R^{ab}R^{cd}e^{e}-\frac{1}{5\ell^{5}}e^{a}e^{b}e^{c}e^{d}e^{e}%
\right) \,,
\end{align}
where $\sigma_{1}$ is a dimensionless arbitrary constant and the factor $k$
carries suitable dimensions making the functional to have units of an
action. Comparing this with Eq.~(\ref{PL1a}) written as
\begin{align}
I^{5D}_{\text{PL}} & =\ell\kappa\int\epsilon_{abcde}\left( \frac{1}{\ell}%
R^{ab}R^{cd}e^{e}-\frac{1}{5\ell^{5}}e^{a}e^{b}e^{c}e^{d}e^{e}\right) \,,
\label{PL2b}
\end{align}
we immediately recognize that
\begin{align}
k=\sigma_{1}^{-1}\ell\kappa\,,
\end{align}
which is clearly dimensionless.

The same line of reasoning was performed in Section 4 to analyze other
examples of $\mathfrak{C}_{m}$ in higher odd dimensions.

\section{Appendix\label{App2}}

In this appendix, we present the explicit expression of the $5D$ $\mathfrak{C%
}_{7}$-CS action. To this purpose, let us consider the explicit connection
one-form%
\begin{equation}
A=\frac{1}{2}\omega ^{ab}J_{ab}+\frac{1}{l}e^{a}P_{a}+\frac{1}{2}%
k^{ab,\left( 1\right) }Z_{ab,\left( 1\right) }+\frac{1}{l}h^{a,\left(
1\right) }R_{a,\left( 1\right) }+\frac{1}{2}k^{ab,\left( 2\right)
}Z_{ab,\left( 2\right) }+\frac{1}{l}h^{a,\left( 2\right) }Z_{a,\left(
2\right) }\,,  \label{1fc7}
\end{equation}%
and the $18$ non-vanishing components of the invariant tensor for the $%
\mathfrak{C}_{7}$ algebra%
\begin{equation}
\left\langle J_{ab,\left( q\right) }J_{cd,\left( r\right) }P_{e,\left(
s\right) }\right\rangle =\frac{4}{3}\sigma _{2u+1}\delta _{j\left(
q,r,s\right) }^{u}\epsilon _{abcde}\,.
\end{equation}%
After using the connection one-form (\ref{1fc7}) and the non-vanishing
components of the invariant tensor in the general expression for the $5D$ CS
action (\ref{actm}) we obtain%
\begin{align}
I_{\mathfrak{C}_{7}\text{-CS}}^{5D}=& k\int \sigma _{1}\left[ \epsilon
_{abcde}\frac{1}{\ell }R^{ab}R^{cd}e^{e}+\mathcal{\tilde{L}}_{1}\left(
\omega ,e,k^{\left( 1\right) },h^{\left( 1\right) },k^{\left( 2\right)
},h^{\left( 2\right) }\right) \right]  \notag \\
& +\sigma _{3}\left[ \epsilon _{abcde}\frac{2}{3\ell ^{3}}%
R^{ab}e^{c}e^{d}e^{e}+\mathcal{\tilde{L}}_{3}\left( \omega ,e,k^{\left(
1\right) },h^{\left( 1\right) },k^{\left( 2\right) },h^{\left( 2\right)
}\right) \right] \,  \notag \\
& +\sigma _{5}\left[ \epsilon _{abcde}\frac{1}{5\ell ^{5}}%
e^{a}e^{b}e^{c}e^{d}e^{e}+\mathcal{\tilde{L}}_{5}\left( \omega ,e,k^{\left(
1\right) },h^{\left( 1\right) },k^{\left( 2\right) },h^{\left( 2\right)
}\right) \right] ,
\end{align}%
where $\mathcal{\tilde{L}}_{1}$, $\mathcal{\tilde{L}}_{3}$, $\mathcal{\tilde{%
L}}_{5}$ are explicitly given by
\begin{align*}
\mathcal{\tilde{L}}_{1}& =\epsilon _{abcde}\left( \frac{4}{\ell }R^{ab}k_{%
\text{ }f}^{c,\left( 1\right) }k^{fd,\left( 2\right) }+\frac{4}{\ell }k_{%
\text{ }f}^{a,\left( 1\right) }k^{fb,\left( 2\right) }k_{\text{ }%
g}^{c,\left( 1\right) }k^{gd,\left( 2\right) }+\frac{2}{\ell }R^{ab,\left(
1\right) }R^{cd,\left( 2\right) }+\frac{2}{\ell ^{3}}R^{ab}h^{c,\left(
1\right) }h^{d,\left( 1\right) }\right. \\
& \left. +\frac{2}{\ell ^{3}}R^{ab,\left( 1\right) }h^{c,\left( 1\right)
}e^{d}+\frac{2}{\ell ^{3}}R^{ab,\left( 1\right) }h^{c,\left( 2\right)
}h^{d,\left( 2\right) }+\frac{2}{3\ell ^{3}}R^{ab,\left( 2\right)
}e^{c}e^{d}+\frac{1}{\ell ^{5}}h^{a,\left( 1\right) }e^{b}e^{c}e^{d}\right.
\\
& \left. +\frac{1}{\ell ^{5}}h^{a,\left( 1\right) }h^{b,\left( 1\right)
}h^{c,\left( 1\right) }h^{d,\left( 1\right) }+\frac{2}{\ell ^{5}}h^{a,\left(
2\right) }h^{b,\left( 2\right) }e^{c}e^{d}+\frac{6}{\ell ^{5}}h^{a,\left(
1\right) }h^{b,\left( 1\right) }h^{c,\left( 2\right) }e^{d}\right. \\
& \left. +\frac{4}{\ell ^{5}}h^{a,\left( 1\right) }h^{b,\left( 2\right)
}h^{c,\left( 2\right) }h^{d,\left( 2\right) }\right) e^{e}+\epsilon
_{abcde}\left( \frac{2}{\ell }R^{ab}R^{cd,\left( 2\right) }+\frac{1}{\ell }%
R^{ab,\left( 1\right) }R^{cd,\left( 1\right) }+\frac{4}{\ell }R^{ab,\left(
2\right) }k_{\text{ }g}^{c,\left( 1\right) }k^{gd,\left( 2\right) }\right. \\
& \left. +\frac{2}{3\ell ^{3}}R^{ab,\left( 2\right) }h^{c,\left( 1\right)
}h^{d,\left( 1\right) }+\frac{6}{\ell ^{3}}R^{ab,\left( 2\right)
}h^{c,\left( 2\right) }e^{d}+\frac{2}{\ell ^{3}}R^{ab,\left( 1\right)
}h^{c,\left( 1\right) }h^{d,\left( 2\right) }+\frac{2}{\ell ^{3}}%
R^{ab}h^{c,\left( 2\right) }h^{d,\left( 2\right) }\right. \\
& \left. +\frac{2}{\ell ^{5}}h^{a,\left( 2\right) }h^{b,\left( 2\right)
}h^{c,\left( 1\right) }h^{d,\left( 1\right) }\right) h^{e,\left( 1\right)
}+\epsilon _{abcde}\left( \frac{2}{\ell }R^{ab}R^{cd,\left( 1\right) }+\frac{%
1}{\ell }R^{ab,\left( 2\right) }R^{cd,\left( 2\right) }+\frac{4}{\ell }%
R^{ab,\left( 1\right) }k_{\text{ }g}^{c,\left( 1\right) }k^{gd,\left(
2\right) }\right. \\
& \left. +\frac{2}{\ell ^{3}}R^{ab}e^{c}e^{d}+\frac{2}{\ell ^{3}}%
R^{ab}h^{c,\left( 1\right) }h^{d,\left( 2\right) }+\frac{2}{\ell ^{3}}%
R^{ab,\left( 1\right) }h^{c,\left( 1\right) }h^{d,\left( 1\right) }+\frac{6}{%
\ell ^{3}}R^{ab,\left( 2\right) }h^{c,\left( 1\right) }e^{d}\right. \\
& \left. +\frac{2}{3\ell ^{3}}R^{ab,\left( 2\right) }h^{c,\left( 2\right)
}h^{d,\left( 2\right) }+\frac{1}{5\ell ^{5}}h^{a,\left( 2\right)
}h^{b,\left( 2\right) }h^{c,\left( 2\right) }h^{d,\left( 2\right) }\right)
h^{e,\left( 2\right) }\,,
\end{align*}%
\begin{align*}
\mathcal{\tilde{L}}_{3}& =\epsilon _{abcde}\left( \frac{2}{\ell }%
R^{ab}R^{cd,\left( 1\right) }+\frac{1}{\ell }R^{ab,\left( 2\right)
}R^{cd,\left( 2\right) }+\frac{4}{\ell }R^{ab,\left( 1\right) }k_{\text{ }%
g}^{c,\left( 1\right) }k^{gd,\left( 2\right) }+\frac{6}{\ell ^{3}}%
R^{ab}h^{c,\left( 1\right) }h^{d,\left( 2\right) }\right. \\
& \left. +\frac{2}{\ell ^{3}}R^{ab,\left( 1\right) }h^{c,\left( 1\right)
}h^{d,\left( 1\right) }+\frac{2}{\ell ^{3}}R^{ab,\left( 1\right)
}h^{c,\left( 2\right) }e^{d}+\frac{2}{\ell ^{3}}R^{ab,\left( 2\right)
}h^{c,\left( 1\right) }e^{d}+\frac{2}{\ell ^{3}}R^{ab,\left( 2\right)
}h^{c,\left( 2\right) }h^{d,\left( 2\right) }\right. \\
& \left. +\frac{2}{\ell ^{5}}h^{a,\left( 1\right) }h^{b,\left( 1\right)
}e^{c}e^{d}+\frac{1}{\ell ^{5}}h^{a,\left( 2\right) }e^{b}e^{c}e^{d}+\frac{1%
}{\ell ^{5}}h^{a,\left( 2\right) }h^{b,\left( 2\right) }h^{c,\left( 2\right)
}h^{d,\left( 2\right) }+\frac{4}{\ell ^{5}}h^{a,\left( 1\right) }h^{b,\left(
2\right) }h^{c,\left( 1\right) }h^{d,\left( 1\right) }\right. \\
& \left. +\frac{6}{\ell ^{5}}h^{a,\left( 1\right) }h^{b,\left( 2\right)
}h^{c,\left( 2\right) }e^{d}\right) e^{e}+\epsilon _{abcde}\left( \frac{1}{%
\ell }R^{ab}R^{cd}+\frac{2}{\ell }R^{ab,\left( 1\right) }R^{cd,\left(
2\right) }+\frac{4}{\ell }R^{ab}k_{\text{ }g}^{c,\left( 1\right)
}k^{gd,\left( 2\right) }\right. \\
& \left. +\frac{2}{3\ell ^{3}}R^{ab}h^{c,\left( 1\right) }h^{d,\left(
1\right) }+\frac{2}{\ell ^{3}}R^{ab,\left( 1\right) }h^{c,\left( 2\right)
}h^{d,\left( 2\right) }+\frac{2}{\ell ^{3}}R^{ab,\left( 2\right)
}h^{c,\left( 1\right) }h^{d,\left( 2\right) }\right. \\
& \left. +\frac{1}{5\ell ^{5}}h^{a,\left( 1\right) }h^{b,\left( 1\right)
}h^{c,\left( 1\right) }h^{d,\left( 1\right) }+\frac{2}{\ell ^{5}}h^{a,\left(
2\right) }h^{b,\left( 2\right) }h^{c,\left( 2\right) }h^{d,\left( 1\right)
}\right) h^{e,\left( 1\right) } \\
& +\epsilon _{abcde}\left( \frac{2}{\ell }R^{ab}R^{cd,\left( 2\right) }+%
\frac{1}{\ell }R^{ab,\left( 1\right) }R^{cd,\left( 1\right) }+\frac{4}{\ell }%
R^{ab,\left( 2\right) }k_{\text{ }g}^{c,\left( 1\right) }k^{gd,\left(
2\right) }+\frac{2}{\ell ^{3}}R^{ab,\left( 2\right) }h^{c,\left( 1\right)
}h^{d,\left( 1\right) }\right. \\
& \left. +\frac{2}{\ell ^{3}}R^{ab,\left( 1\right) }h^{c,\left( 1\right)
}h^{d,\left( 2\right) }+\frac{2}{3\ell ^{3}}R^{ab}h^{c,\left( 2\right)
}h^{d,\left( 2\right) }\right) h^{e,\left( 2\right) }\,,
\end{align*}%
and%
\begin{align*}
\mathcal{\tilde{L}}_{5}& =\epsilon _{abcde}\left( \frac{2}{\ell }%
R^{ab}R^{cd,\left( 2\right) }+\frac{1}{\ell }R^{ab,\left( 1\right)
}R^{cd,\left( 1\right) }+\frac{4}{\ell }R^{ab,\left( 2\right) }k_{\text{ }%
g}^{c,\left( 1\right) }k^{gd,\left( 2\right) }+\frac{2}{\ell ^{3}}%
R^{ab}h^{c,\left( 2\right) }h^{d,\left( 2\right) }\right. \\
& \left. +\frac{2}{3\ell ^{3}}R^{ab,\left( 1\right) }e^{c}e^{d}+\frac{2}{%
\ell ^{3}}R^{ab,\left( 2\right) }h^{c,\left( 1\right) }h^{d,\left( 1\right)
}+\frac{2}{\ell ^{3}}R^{ab,\left( 2\right) }h^{c,\left( 2\right) }e^{d}+%
\frac{2}{\ell ^{5}}h^{a,\left( 1\right) }h^{b,\left( 1\right) }h^{c,\left(
1\right) }e^{d}\right. \\
& \left. +\frac{2}{\ell ^{5}}h^{a,\left( 2\right) }h^{b,\left( 2\right)
}h^{c,\left( 2\right) }e^{d}+\frac{4}{\ell ^{5}}h^{a,\left( 1\right)
}h^{b,\left( 2\right) }e^{c}e^{d}+\frac{6}{\ell ^{5}}h^{a,\left( 1\right)
}h^{b,\left( 2\right) }h^{c,\left( 2\right) }h^{d,\left( 1\right) }\right)
e^{e} \\
& +\epsilon _{abcde}\left( \frac{2}{\ell }R^{ab}R^{cd,\left( 1\right) }+%
\frac{1}{\ell }R^{ab,\left( 2\right) }R^{cd,\left( 2\right) }+\frac{4}{\ell }%
R^{ab,\left( 1\right) }k_{\text{ }g}^{c,\left( 1\right) }k^{gd,\left(
2\right) }+\frac{2}{\ell ^{3}}R^{ab}e^{c}e^{d}\right. \\
& \left. +\frac{2}{\ell ^{3}}R^{ab}h^{c,\left( 1\right) }h^{d,\left(
2\right) }+\frac{6}{\ell ^{3}}R^{ab,\left( 1\right) }h^{c,\left( 2\right)
}e^{d}+\frac{2}{3\ell ^{3}}R^{ab,\left( 1\right) }h^{c,\left( 1\right)
}h^{d,\left( 1\right) }+\frac{2}{\ell ^{3}}R^{ab,\left( 2\right)
}h^{c,\left( 2\right) }h^{d,\left( 2\right) }\right. \\
& \left. +\frac{1}{\ell ^{5}}h^{a,\left( 2\right) }h^{b,\left( 1\right)
}h^{c,\left( 1\right) }h^{d,\left( 1\right) }+\frac{1}{\ell ^{5}}h^{a,\left(
2\right) }h^{b,\left( 2\right) }h^{c,\left( 2\right) }h^{d,\left( 2\right)
}\right) h^{e,\left( 1\right) } \\
& +\epsilon _{abcde}\left( \frac{1}{\ell }R^{ab}R^{cd}+\frac{2}{\ell }%
R^{ab,\left( 1\right) }R^{cd,\left( 2\right) }+\frac{4}{\ell }R^{ab}k_{\text{
}g}^{c,\left( 1\right) }k^{gd,\left( 2\right) }+\frac{2}{\ell ^{3}}%
R^{ab}h^{c,\left( 1\right) }h^{d,\left( 1\right) }\right. \\
& \left. +\frac{6}{\ell ^{3}}R^{ab,\left( 1\right) }h^{c,\left( 1\right)
}e^{d}+\frac{2}{3\ell ^{3}}R^{ab,\left( 1\right) }h^{c,\left( 2\right)
}h^{d,\left( 2\right) }+\frac{2}{\ell ^{3}}R^{ab,\left( 2\right)
}h^{c,\left( 1\right) }h^{d,\left( 2\right) }\right) h^{e,\left( 2\right)
}\,,
\end{align*}%
where we have defined%
\begin{align*}
R^{ab,\left( 1\right) }& =Dk^{ab,\left( 1\right) }+k_{\text{ }c}^{a,\left(
2\right) }k^{cb,\left( 2\right) }\,, \\
R^{ab,\left( 2\right) }& =Dk^{ab,\left( 2\right) }+k_{\text{ }c}^{a,\left(
1\right) }k^{cb,\left( 1\right) }\,.
\end{align*}

\end{document}